\documentclass[12pt,a4paper]{article}
\usepackage{amsmath,amsfonts,amssymb,amsthm}
\usepackage[T1]{fontenc}
\usepackage{hyperref}
\usepackage{graphicx}
\usepackage{stmaryrd}
\usepackage{cite}

\newcommand{\be}{\begin{equation}}
\newcommand{\ee}{\end{equation}}
\newcommand{\bea}{\begin{eqnarray}}
\newcommand{\eea}{\end{eqnarray}}

\newcommand{\phii}{\phi}
\newcommand{\BMS}{\mathrm{BMS}_3}

\newcommand{\Diff}{\mathrm{Diff}^+(S^1)}

\newcommand{\Vect}{\mathrm{Vect}(S^1)}

\newcommand{\hatVect}{\widehat{\mathrm{Vect}}(S^1)}
\newcommand{\hatDiff}{\widehat{\mathrm{Diff}^+}(S^1)}
\newcommand{\hatbms}{\widehat{\mathfrak{bms}}_3}
\newcommand{\hatBMS}{\widehat{\mathrm{BMS}}_3}

\newcommand{\Ad}{\mathrm{Ad}}

\newcommand{\RR}{\mathbb{R}}

\newcommand{\NN}{\mathbb{N}}

\def\d_Vphi{\text{d}_V\hspace{-0.06em}\phi}
\def\d_Vphibar{\text{d}_V\hspace{-0.06em}\bar\phi}
\def\d_Vxi{\text{d}_V\hspace{-0.06em}\xi}

\def\be{\begin{eqnarray}}
\def\ee{\end{eqnarray}}
\def\beann{\begin{eqnarray*}}
\def\eeann{\end{eqnarray*}}
\def\beq{\begin{equation}}
\def\eeq{\end{equation}}
\def\ba{\begin{array}}
\def\ea{\end{array}}
\def\ben{\begin{enumerate}}
\def\een{\end{enumerate}}
\def\bea{\begin{eqnarray}}
\def\eea{\end{eqnarray}}

\def\5{\bar }
\def\6{\partial }
\def\7{\hat }
\def\4{\tilde }
\advance\textheight\topskip
\renewcommand{\tilde}{\widetilde}
\renewcommand{\hat}{\widehat}

\renewcommand{\d}{\partial}

\renewcommand{\geq}{\,{\geqslant}\,}
\renewcommand{\leq}{\,{\leqslant}\,}

\newcommand{\binner}[2]{%
  {\langle}\kern-4.15pt{\langle}#1{,}\,#2{\rangle}\kern-4.15pt{\rangle}}

\newcommand{\half}{\frac{1}{2}}

\newcommand{\ffrac}[2]{\raisebox{.5pt}%
  {\footnotesize$\displaystyle\frac{#1}{#2}$}\kern1pt}

\newcommand{\ZZ}{\mathbb{Z}}

\numberwithin{equation}{section} \makeatletter
\@addtoreset{equation}{section}

\DeclareFontFamily{OT1}{rsfs}{} \DeclareFontShape{OT1}{rsfs}{m}{n}{
<-7> rsfs5 <7-10> rsfs7 <10-> rsfs10}{}
\DeclareMathAlphabet{\mycal}{OT1}{rsfs}{m}{n}

\begin{document}

\title{Holographic positive energy theorems in three-dimensional gravity}

\author{Glenn Barnich and Blagoje Oblak}

\date{}

\def\mytitle{Holographic positive energy theorems \\ in three-dimensional
  gravity}

\pagestyle{myheadings} \markboth{\textsc{\small G.~Barnich,
    B.~Oblak}}{%
  \textsc{\small Positive energy in 3d gravity}}

\addtolength{\headsep}{4pt}


\begin{centering}

  \vspace{1cm}

  \textbf{\Large{\mytitle}}


  \vspace{1.5cm}

  {\large Glenn Barnich$^{a}$ and Blagoje Oblak$^{b}$}

\vspace{.5cm}

\begin{minipage}{.9\textwidth}\small \it  \begin{center}
   Physique Th\'eorique et Math\'ematique \\ Universit\'e Libre de
   Bruxelles and International Solvay Institutes \\ Campus
   Plaine C.P. 231, B-1050 Bruxelles, Belgium
 \end{center}
\end{minipage}

\end{centering}

\vspace{1cm}

\begin{center}
  \begin{minipage}{.9\textwidth}
    \textsc{Abstract}. The covariant phase space of three-dimensional
    asymptotically flat and anti-de Sitter gravity is controlled by
    well-understood coadjoint orbits of the Virasoro group. Detailed
    knowledge on the behavior of the energy functional on these orbits
    can be used to discuss positive energy theorems.
  \end{minipage}
\end{center}

\vfill

\noindent
\mbox{}
\raisebox{-3\baselineskip}{%
  \parbox{\textwidth}{\mbox{}\hrulefill\\[-4pt]}}
{\scriptsize$^a$Research Director of the Fund for Scientific
  Research-FNRS Belgium. E-mail: gbarnich@ulb.ac.be\\
$^b$ Research Fellow of the Fund for Scientific Research-FNRS
Belgium. E-mail: boblak@ulb.ac.be}

\thispagestyle{empty}
\newpage


\section{Introduction}
\label{sec:introduction}

The nature of energy-momentum in general relativity is more subtle
than for standard field theories on Minkowski spacetime because
translations are local, rather than global, symmetries. As a
consequence, the definition of energy-momentum in terms of the metric
requires suitable asymptotic conditions and involves integrals over
surfaces of codimension two rather than one. A natural question is
then to understand under which assumptions the classical energy is
bounded from below, and if it is Minkowski spacetime that minimizes the
energy.

Much work has been devoted to settle these questions in four
dimensions in the asymptotically flat case, both at spatial infinity
for the ADM mass \cite{Arnowitt:1960zzc,Arnowitt:1962aa}, and at null
infinity for the Bondi mass \cite{Bondi:1962px,Sachs1962a}. A clear
formulation of the problem and the state of the art until 1979 is
described in \cite{Brill1980}. For the ADM mass, this problem has been
solved in \cite{Schon:1981vd}, with a simplified proof given in
\cite{Witten:1981mf} using supersymmetry-motivated arguments (see also
\cite{Deser:1977hu,Grisaru:1977gj} for earlier work and
\cite{Parker:1981uy,Nester1981,Reula:1982xt} for refinements); the
case of the Bondi mass has been treated in
\cite{Israel1981,Ludvigsen:1981gf,Schon:1982re,%
  Horowitz:1981uw,Ashtekar1982a,Reula1984}.

Gravity in three dimensions has turned out to be a surprisingly rich
toy model for several aspects of four-dimensional general
relativity. When solution space is restricted to conical defects and
excesses \cite{Deser:1983tn,Deser:1983dr} or other zero-mode solutions
such as the BTZ black holes \cite{Banados:1992wn,Banados:1993gq},
positivity of energy can be addressed directly from the detailed
understanding of these solutions. Upon allowing for non-trivial
asymptotics however, solution space becomes infinite-dimensional and
positivity of energy is a real issue, even though there are no bulk
gravitons. The purpose of this letter is to point out that, in this
setting, positive energy theorems in three dimensions remain
explicitly tractable and can be discussed in terms of non-trivial
properties of coadjoint orbits of the Virasoro group, both in the
asymptotically flat and in the asymptotically anti-de Sitter cases.

More precisely, our reasoning is the following:

(i) The covariant phase space of three-dimensional gravity turns out
to coincide with the subspace at fixed central charges of the
coadjoint representation of the relevant asymptotic symmetry group. In
the asymptotically flat case, this is the centrally extended BMS$_3$
group, while in the anti-de Sitter case it is the direct product of
two Virasoro groups. In both cases, the mass is expressed in terms of
the energy functional on coadjoint orbits of the Virasoro group.

(ii) The complete classification of the Virasoro coadjoint orbits is
a classical result \cite{LazPan,Segal:1981ap}, discussed in many
places in the literature.

(iii) The behavior of the energy functional on each Virasoro orbit is
explicitly known \cite{Witten:1987ty,Balog:1997zz}. According to a
footnote, an argument in \cite{Witten:1987ty} aiming at the
understanding of the Virasoro energy functional is patterned on work
done in the context of the positive energy conjecture in
four-dimensional gravity \cite{Brill1968}. In turn, we show here that
the analysis of the energy functional on Virasoro coadjoint orbits, as
completed in \cite{Balog:1997zz}, allows one to address positive
energy theorems in three dimensions.

(iv) In the asymptotically flat case, this analysis allows us to
readily isolate the orbits that have to be discarded in order for the
Bondi mass to be bounded from below, leaving only solutions whose mass
is greater or equal to that of Minkowski spacetime. The discussion in
the asymptotically anti-de Sitter case, both for the Bondi and the ADM
mass, is more subtle and relies on the details of how the two chiral
sectors of the boundary theory should be combined.

In the context of a purely classical and gravitational version of
holography, our approach can be given the following
interpretation. The Chern-Simons formulation
\cite{Achucarro:1987vz,Witten:1988hc} of three-dimensional gravity
allows one to implement, on the level of action principles, the
equivalence (in the presence of boundaries or non trivial boundary
conditions) with a two-dimensional Wess-Zumino-Witten theory
\cite{Elitzur:1989nr}. When taking all gravitational boundary
conditions into account, a further reduction to Liouville theory is
implemented in the AdS$_3$ case \cite{Coussaert:1995zp}, and to a flat
limit thereof in the flat case \cite{Barnich:2013yka}. In this
context, the gravitational energy functionals that we have have been
studying here are the energy functionals associated with the global
time translation symmetry of these two-dimensional field theories.

\section{Covariant phase space of 3d flat gravity}
\label{sec:covar-phase-space-1}

As in the better known anti-de Sitter case, three-dimensional
asymptotically flat spacetimes at null infinity are entirely
determined by their symmetry structure.

\subsubsection*{BMS$_{\bf 3}$ group}

The relevant asymptotic symmetry group \cite{Ashtekar1997} is the
semi-direct product
\begin{equation}
  \label{eq:20}
  \BMS:=\Diff\ltimes_{\Ad}\Vect.
\end{equation}
Here $\Diff$ is the group of orientation-preserving diffeomorphisms
of the circle, called {\it superrotations}. It generalizes the Lorentz
subgroup of the three-dimensional Poincar\'e group. On the other hand,
$\Vect$ is the abelian additive group of vector fields on the circle,
which can be identified with densities of weight $-1$. They are called
{\it supertranslations} in this context, and generalize Poincar\'e
translations. Superrotations, denoted by $f$, act on
supertranslations, denoted by $\alpha$, according to the adjoint
action denoted by $\cdot$ below.

Points of the circle will be labelled using an angular coordinate
$\phii$ identified as $\phii\sim\phii+2\pi$. Functions on the circle
may then be seen as $2\pi$-periodic functions of $\phii$. Similarly,
diffeomorphisms of the circle can be seen as diffeomorphisms $f$ of
$\RR$ that satisfy $f(\phii+2\pi)=f(\phii)+2\pi$ and
$f'(\phii)>0$. This amounts to going to the universal cover of
$\Diff$.

Representations in terms of surface charges rely on the central
extension $\hatBMS$ $=\hatDiff\ltimes_{\Ad}\hatVect$. Here, the first
factor is the Virasoro group, while the second one is its algebra,
seen as an abelian additive group. The associated Lie algebra, denoted
by $\hatbms$, is the semi-direct sum of the Virasoro algebra with its
adjoint representation. Its elements are of the form $(y,-ia;t,-ib)$,
where $y=Y(\phii)\d/{\d\phii}$ is a vector field on the circle,
$t=T(\phii)d\phii^{-1}$ is a density of weight $-1$, and $a,b$ are
real numbers.

\subsubsection*{Coadjoint representation of BMS$_{\bf 3}$}

Elements $(j,ic_1;p,ic_2)$ of the coadjoint representation $\hatbms^*$
contain, besides the central charges $c_1,c_2\in \RR$, the quadratic
differentials $j=J(\phii)d\phii^2$ and $p=P(\phii)d\phii^2$. It is
tempting to call them angular and linear {\it supermomentum}, respectively,
because they can be interpreted as infinite-dimensional
generalizations of angular and linear momentum. Their pairing with
elements of $\hatbms$ is given by
\begin{equation}
  \label{eq:37}
  \langle (j,ic_1;p,ic_2), (y,-ia;t,-ib)\rangle
  =\int^{2\pi}_0d\phii\, 
  (JY+PT)+ c_1a
  + c_2b.
\end{equation}
The coadjoint action can then readily be worked out and reads
\begin{equation}
   \Ad^*_{(f^{-1};-f^{-1}\cdot\alpha)}(j,ic_1;p,ic_2)=(\tilde j,i
   c_1;\tilde p,i c_2),  \label{eq:1}
\end{equation}
with 
\begin{eqnarray}
  \tilde P & = & P\circ f (f')^2-
  \frac{c_2}{24\pi}S[f],\label{en}\\
  \tilde J & = & \left[J+\alpha
    P'+2\alpha'P-\frac{c_2}{24\pi}\alpha'''\right]\circ f  
  (f')^2-\frac{c_1}{24\pi} S[f],\nonumber
\end{eqnarray}
where $S[f]= {f'''}/{f'}-{3}/{2}({f''}/{f'})^2$ is the Schwarzian
derivative of $f$.

\subsubsection*{Energy functional}

The energy associated with a coadjoint vector is conjugate to a time
translation and is defined to be the zero mode of $P(\phii)$,
\begin{equation}
  \label{eq:70}
  E_P:=\int^{2\pi}_0d\phii\, P.
\end{equation}
Similarly, in gravity, the lowest Fourier modes $P_{\pm 1}$ are conjugate to
spatial translations and encode linear momentum.

\subsubsection*{Asymptotically flat gravity}
 
The reduced phase space of three-dimensional asymptotically flat
gravity with its Dirac bracket
\cite{Barnich:2006avcorr,Barnich:2010eb} can be identified with
the subspace of $\hatbms^*$ at fixed central charges
\begin{equation}
c_1=0,\quad c_2=\frac{3}{G},\label{eq:28}
\end{equation}
equipped with the Kirillov-Kostant Poisson bracket. Indeed, after a
suitable gauge fixing, the general solution to Einstein's equations
describing three-dimensional asymptotically flat spacetimes at null
infinity is given by metrics
\begin{equation}
  \label{eq:2}
  ds^2=\Theta du^2-2dudr+2\left(\Xi+\frac{u}{2}\Theta'\right)dud\phii+r^2d\phii^2,
\end{equation}
depending on two arbitrary functions
$\Theta=\Theta(\phii),\Xi=\Xi(\phii)$. Under finite asymptotic
symmetry transformations, the latter have been shown
\cite{Barnich:2012rz} to transform as in (\ref{eq:1}) after
identifying $\Theta= (16\pi G) P$, $\Xi = (8\pi G) J$, and switching
from a passive to an active point of view. The associated surface
charges, normalized with respect to the null orbifold $\Theta=0=\Xi$,
coincide with \eqref{eq:37} up to central terms. In particular, the
Bondi energy is given by \eqref{eq:70}.

As a consequence, in the asymptotically flat case, the gravitational
solution space is classified by coadjoint orbits of $\hatBMS$. On
account of the semi-direct product structure, these orbits are
determined by the knowledge of the coadjoint orbits of the Virasoro
group. Furthermore it follows from (\ref{en}) that, for questions
concerning the behavior of the energy functional (\ref{eq:70}), the
complete classification of the coadjoint orbits of $\hatBMS$ is not
actually needed: the classification of those of the Virasoro group (at
central charge $c_2$) is enough.

\section{Energy bounds in flat gravity}
\label{sec:coadj-orbits-viras}

\subsubsection*{Coadjoint orbits of the Virasoro group} 

Since the coadjoint representation of the Virasoro group is described
by elements $(p,ic_2)$ transforming as in (\ref{en}), the little group
of such an element consists of diffeomorphisms $f$ such that $\tilde
P=P$.

In the gravitational context $c_2={3}/{G}> 0$, so we focus below on
the case of non-vanishing, strictly positive central charge. Then the
simplest orbits are those that admit a constant representative
$P(\phii)=k$. For generic $k$, the little group is the group ${\rm
  U}(1)$ of rigid rotations of the circle. The only exceptions are the
values $k=-{c_2 n^2}/{48\pi}$, $n\in \NN^*$, for which the little
group is the $n$-fold cover ${\rm PSL}^{(n)}(2,\RR)$ of ${\rm
  PSL}(2,\RR)$. It consists of diffeomorphisms $f_n$ of the form
    \begin{equation}
      \label{eq:53} e^{in f_n(\phi)}=\frac{\alpha
e^{in\phii}+\beta}{\bar\beta
e^{in\phii}+\bar\alpha},\quad |\alpha|^2-|\beta|^2=1.
\end{equation} 

There are two additional families of Virasoro orbits, containing no
constant coadjoint vectors.

(i) The orbits of the first family are labelled by parameters $\mu^2>0$
and $n\in \NN^*$. Each such orbit can be understood as a tachyonic
deformation of the orbit of the exceptional constant $-{c_2
  n^2}/{48\pi}$, which is recovered in the limit $\mu\to 0$. The little
group is isomorphic to $\RR^*_+\times \ZZ_n$, where $\RR^*_+$ is the
multiplicative group of strictly positive real numbers and $\ZZ_n$
acts on the circle by rigid rotations by multiples of the
angle $2\pi/n$.

(ii) The orbits of the second family are characterized by $q\in \{\pm
1\}$ and $n\in \NN^*$, with the same little groups as in the previous
case.  Such orbits can be understood as future- or past-directed
massless deformations of the orbits of the exceptional constants.

\subsubsection*{Energy bounds on Virasoro coadjoint orbits}

The energy on a Virasoro coadjoint orbit is
given by
\begin{equation}
  \label{eq:63a}
  E_{P}[f]:=\int_{0}^{2\pi}d\phii\,\left[\theta
    P+\frac{c_2}{48\pi}\frac{(\theta')^2}{\theta}\right], 
\end{equation}
with $\theta:=f'\circ f^{-1}$. The last term can be made arbitrarily
large so that the energy is unbounded from above on every orbit.

Now, an important property of the Schwarzian derivative is the
Schwartz inequality \cite{Schwartz1992} (see also \cite{Balog:1997zz}
for an elementary proof)
\begin{equation}
  \label{eq:64}
  \int_0^{2\pi} d\phii\, \left[S[f]+\half ((f')^2-1)\right]
  \leq 0\quad\forall\, f\in\Diff,
\end{equation}
with equality iff $f$ is the lift of a projective transformation of
the circle, that is, a transformation of the form (\ref{eq:53}) with
$n=1$. This inequality readily implies that the energy is bounded from
below on the orbit of the constant $-{c_2}/{48\pi}$, and that this
constant realizes the minimum value of energy, $-c_2/24$.

More generally, for an arbitrary constant $P=k\in \RR$, 
\begin{equation}
  \label{eq:71}
  E_k[f]
  =
  \left(E_{-\frac{c_2}{48\pi}}[f]+\frac{c_2}{24}\right)
  +2\pi k
  +\left(k+\frac{c_2}{48\pi}\right)\int^{2\pi}_0d\phii\,(f'-1)^2,
\end{equation}
so that the energy is manifestly bounded from below on the orbit of
$P=k$ iff $k\geq -{c_2}/{48\pi}$, the global
minimum being reached at $P=k$ itself, with minimal energy $2\pi k$.

Finally, one can show \cite{Balog:1997zz} that the energy is unbounded
from below on all orbits containing no constant representatives, except for
the massless deformation of $k=-{c_2}/{48\pi}$ with $q=-1$. In this
case the lower bound of energy is again $-{c_2}/{24}$, but it is
not reached on the orbit.

\subsubsection*{Application to flat gravity} 

In order to have energy bounded from below, all solutions belonging to
the orbits of constant coadjoint vectors $k<-{c_2}/{48\pi}$ must be
discarded. Solutions belonging to the orbits without constant
representatives must be discarded as well, except for those in the
orbit of the massless deformation of $k=-{c_2}/{48\pi}$ with
$q=-1$. When this is done, the absolute minimum of the energy is
$-{c_2}/{24}=-{1}/{8G}$, which is reached at $P=-{c_2}/{48\pi}$.  The
associated metrics are given by (\ref{eq:2}) with $\Theta=-1$ and
arbitrary $\Xi(\phii)$, and contain Minkowski spacetime for
$\Xi=0$. The set of allowed solutions includes, besides the orbit of
the massless deformation of $P=-{c_2}/{48\pi}$ with $q=-1$, the orbits
of all angular defects \cite{Deser:1983tn} and of cosmological
solutions
\cite{Ezawa:1992nk,Cornalba:2002fi,Cornalba:2003kd,Barnich:2012aw},
but not those of angular excesses.

\section{3d AdS gravity and energy bounds}
\label{sec:remark-covar-phase}

The covariant phase space of three-dimensional asymptotically AdS
gravity (with cosmological constant $\Lambda=-\ell^{-2}$) consists of
the subspace, at fixed central charges $c^\pm={3\ell}/{2G}$, of two
disjoint copies $(p^\pm,ic^\pm)$ of the coadjoint representation of
the Virasoro group
\cite{Brown:1986nw,Strominger:1998eq,NavarroSalas:1998ks,%
  Banados1999,Skenderis:1999nb,Navarro-Salas1999,Nakatsu1999}. The
classification of Virasoro coadjoint orbits then also organises this
solution space into disjoint classes. Concretely, this analysis
answers the question which asymptotically AdS$_3$ solutions are related
to each other by Brown-Henneaux transformations in the bulk. For
instance, the solutions belonging to the orbits without constant
representatives described previously cannot be obtained from the
well-known zero mode solutions by a symmetry transformation. A related
discussion on the role of these solutions in the quantum theory has
appeared in \cite{Maloney:2007ud}, while further details on the
classical level can be found in \cite{Scarinci:2011np}.

The total energy normalized with respect to the $M=0=J$ BTZ black hole
\cite{Banados:1992wn,Banados:1993gq} is given by
$M=(E_{P^{+}}+E_{P^{-}})/\ell$, while the angular momentum is
$J=E_{P^{+}}-E_{P^{-}}$. The requirement that the total energy $M$ be
bounded from below then requires both chiral energies $E_{P^{+}}$ and
$E_{P^{-}}$ to be bounded from below. This forces one to discard all
solutions belonging to the orbit of a constant pair $(k^+,k^-)$ with
either $k^+<-{c^+}/{48\pi}$ or $k^-<-{c^-}/{48\pi}$. Solutions
belonging to orbits without constant representatives must be discarded
too, except if the orbits of the massless deformations of
$k^\pm=-{c^\pm}/{48\pi}$ with $q^\pm=-1$ are involved -- either both of
them, or one of them together with the orbit of a constant
representative above $-{c^\pm}/{48\pi}$.

The absolute minimum of the energy is then attained at the solution 
$P^{\pm}=-{c^\pm}/{48\pi}$ and corresponds to ${\rm AdS_3}$
spacetime with $M=-(c^++c^-)/24\ell=-1/8G$ and $J=0$. More
generally, all allowed solutions have their mass and angular momentum
constrained by $M\geq -{1}/{8G}$ and $|J|\leq
\ell M+\ell/8G$. The allowed solutions include the BTZ black holes,
some of the angular defect solutions (but not all of them) and also
solutions with closed time-like curves and naked singularities.

Note however that the discussion changes when one links the two chiral
sectors of the boundary theory. Suppose for the sake of argument that
the sectors are linked as in global Liouville theory. The results of
section 5 of \cite{Balog:1997zz} then imply that the only orbits with
bounded energy are those of the constants $P^+=P^-=k>0$. In gravity
this would correspond to non-rotating ($J=0$) BTZ black holes, with the
exclusion of the extremal case at $M=0=J$. In that setting, the orbit
of the coadjoint vector $P^{\pm}=-{c^\pm}/{48\pi}$, corresponding to
AdS$_3$ spacetime in gravity, has to be discarded because it
becomes only a local minimum.

Even though to a first approximation the dual theory for
asymptotically ${\rm AdS}_3$ gravity is indeed Liouville theory
\cite{Coussaert:1995zp}, the correct boundary theory for the black
hole sector differs from Liouville theory by zero modes and holonomies
\cite{Henneaux:1999ib}. This means that a more refined analysis is
needed to understand how the chiral sectors should be combined.

\section{Open issues}
\label{sec:issues-questions}

Besides properly taking into account zero modes and holonomies in the
AdS$_3$ case, our approach to positive energy theorems in
three-dimensional gravity should be completed in several directions.

(i) On a technical level, in the asymptotically
flat case, a complete understanding of the coadjoint
orbits of $\hat\BMS$ is necessary in order to investigate the behavior of
the angular momentum functional on these orbits.

(ii) In the asymptotically flat case the discussion above was
concerned only with the Bondi mass. Since the energy is conserved and
since there is no news in three dimensions, the Bondi mass is expected
to be equal to the ADM mass, which for asymptotically flat spacetimes
is defined at spatial infinity. In order to show this explicitly, one
needs to move away from the completely gauge fixed, reduced phase
space point of view that has been adopted here and connect both
asymptotic regimes, as has been done in four dimensions (see
e.g.~\cite{Ashtekar:1980aa} and references therein). Similar remarks
apply to the anti-de Sitter case. Indeed, the discussion carried out
above directly concerns either the ADM mass, when carried out in the
Fefferman-Graham gauge, or what could be called the Bondi mass when
performed in the BMS gauge, since results in both gauges are identical
from a group-theoretical viewpoint \cite{Barnich:2012aw}.

(iii) It would also be interesting to connect our analysis with more
standard approaches to positive energy theorems as applied to three
dimensional gravity, for instance by developing arguments based on
supersymmetry.

(iv) In this work, three-dimensional gravity and the behavior of its
energy functional have been studied as interesting, explicitly
solvable problems in their own right. As a converse to (iii), one
might wonder if and how this approach could be used in four
dimensions. In order to apply such direct methods, a complete control
on solution space is needed. In the asymptotically flat case at null
infinity, this has been achieved in the context of the characteristic
initial value problem in \cite{Sachs1962a,Newman1962a}. One is then
naturally led to the problem of the positivity of the Bondi-Sachs
energy functional, as described in detail in chapter 9.10 of
\cite{Penrose:1986}. Whether group theoretical considerations using
either the globally well-defined or the more recent local version
\cite{Barnich:2010eb} of the BMS$_4$ group can shed a new light in
this context remains to be seen.

More details on technical aspects in three dimensions will be given
elsewhere \cite{Barnich:2014kra,Barnich2014}.

\section{Acknowledgements}
\label{sec:acknowledgements}


This work is supported in part by the Fund for Scientific
Research-FNRS (Belgium), by IISN-Belgium, by ``Communaut\'e fran\c
caise de Belgique - Actions de Recherche
Concert\'ees''. G.B.~acknowledges useful discussions with Marc
Henneaux and Mauricio Leston.



\section{References}

\renewcommand{\section}[2]{}%

\def\cprime{$'$}
\providecommand{\href}[2]{#2}\begingroup\raggedright\endgroup

\end{document}